# Nanoparticle Size Distribution Quantification: Results of a SAXS Inter-Laboratory Comparison

*Brian R. Pauw,\* Claudia Kästner, and Andreas F. Thünemann\**

Federal Institute for Materials Research and Testing (BAM), Unter den Eichen 87, 12205 Berlin, Germany

\*AUTHOR EMAIL and PHONE: andreas.thuenemann@bam.de, +493081041610

ABSTRACT. We present the first world-wide inter-laboratory comparison of small-angle X-ray scattering (SAXS) for nanoparticle sizing. The measurands in this comparison are the mean particle radius, the width of the size distribution and the particle concentration. The investigated sample consists of dispersed silver nanoparticles, surrounded by a stabilizing polymeric shell of poly(acrylic acid). The silver cores dominate the X-ray scattering pattern, leading to the determination of their radii size distribution using i.) Glatter's Indirect Fourier Transformation method, ii) classical model fitting using SASfit and iii) a Monte Carlo fitting approach using McSAS. The application of these three methods to the collected datasets produces consistent mean number- and volume-weighted core radii of $R_n = 2.76$ nm and $R_v = 3.20$ nm, respectively. The corresponding widths of the log-normal radii distribution of the particles were $\sigma_n = 0.65$ nm and $\sigma_v = 0.71$ nm. The particle concentration determined using this method was $3.00 \pm 0.38$ g L$^{-1}$ ($4.20 \pm 0.73 \times 10^{-6}$ mol L$^{-1}$). We show that the results are slightly biased by the choice of data evaluation procedure, but that no substantial differences were found between the results from data measured on a very wide range of instruments: the participating laboratories at synchrotron

SAXS beamlines, commercial and home-made instruments were all able to provide data of high quality. Our results demonstrate that SAXS is a qualified method for revealing particle size distributions in the sub-20 nm region (at least), out of reach for most other analytical methods.

**Introduction**

Nanotechnology is omnipresent in our daily life, and widely considered to be an enabling technology of this century. Recently, however, more critical voices have emerged, asking: "How safe are nanomaterials?"[1] and "Is Nano a Bubble?".[2] Their concerns include the metrological challenges of nanoparticles in "real world samples". The line of reasoning is straightforward: on the one hand we have seen an endless story of new spectacular achievements of the capabilities of characterization of nanoscale structures ranging from 1 to 100 nm (the definition of nano objects according to ISO).[3] Examples include the atomically resolved 3D structure of individual platinum nanocrystals in solution as imaged with electron microscopy,[4] and the elucidation of the 3D-architechture of individual silver nanoparticles by free-electron laser X-ray scattering.[5] On the other hand, finding an appropriate measurement methodology to simply determine whether a given material would fall within the EU nanomaterial definition has been more challenging than expected, with no imminent solution in sight (to our knowledge). Such a methodology would need to determine whether 50% of the number of constituents in a material have a dimension smaller than 100 nm.[6, 7] In particular objects with dimensions between 1 and 20 nm are challenging to count.[8]

Demonstrating that a technique is, in fact, able to reliably elucidate the size distribution and amount of such nano objects is therefore of great importance. To this end, inter-laboratory or "Round-Robin" comparisons - which compare results inferred from measurements of identical samples on different instruments - can demonstrate that reliable results can be obtained irrespective of the utilized instrumentation. Only a few such Round-Robin experiments exist for the analytical methods used in nanotechnology, most notably for single-particle ICP-MS[8, 9] and transmission electron microscopy.[10] Furthermore, only one exists for small-angle neutron scattering (SANS)[11] and none at all for small-angle X-ray scattering (SAXS). SAXS is an uncomplicated bulk nanostructural quantification technique, particularly sensitive to the smaller end of the nanoscale, therefore forming a prime candidate to answer the aforementioned analytical needs. In the absence of a standard methodology, however, a wide range of data collection and correction procedures are being applied in the various laboratories and synchrotrons.[12, 13] The practical effects thereof on the accuracy of the findings have heretofore been poorly understood.

A Round-Robin experiment for SAXS would enable an understanding of its practical precision and accuracy. To this end, a suitable sample of dispersed particles is needed that satisfies particular conditions: dimensions smaller than 10 nm, limited size-dispersity, and with a reasonable scattering contrast and concentration. Such samples were synthesized in our laboratory in the form of poly(acrylic acid) stabilized silver nanoparticles with nominal radii of 3 nm.[14] Silver nanoparticles where chosen since they are one of the most widespread type of nanoparticles in consumer products worldwide and their proper analytics is of high interest.[15] This work provides the first inter-laboratory comparison of the measurement of nanoparticle size distributions with small-angle X-ray scattering (SAXS). The chosen sample consists of silver particles wrapped by poly(acrylic acid) as stabilizer.[14] The measurements received for this sample from the various laboratories are subjected to a trio of fundamentally different analysis methods. The expected outcome is a qualified estimation on how accurate and precise the SAXS method is for determination of sizes of nanoparticles in the sub-20 nm range.

**Experimental Section**

**Participants.**

A total of 45 samples were measured in 22 laboratories (a maximum of two samples per laboratory). Samples were measured from February to May 2016. 19 laboratories measured both (identical) samples, of which two laboratories measured both samples on two different instruments, and one beamline measured one sample at two photon energies. Three laboratories measured one sample only. Many of these laboratories were recruited at the 16[th] International Conference on Small-Angle Scattering in Berlin, while others were recruited via an announcement of the study on a SAXS-related weblog http://www.lookingatnothing.com/. As the purpose of the study is to determine the practical variance between SAXS results, we explicitly refrain from comparing the instruments directly. To that end, all collected datasets have been anonymized thoroughly (and are available in the SI). Each laboratory and user was given a brief instruction set (vide infra), but was otherwise left free to choose their own measurement criteria. The effects of the differences in measurement methodology on the resulting dataset allows us a view on the impact - or lack thereof - on the sizing results.

**Sample Preparation and Measurements**

Nanoparticles were synthesized as published elsewhere.[14] The resulting batch was used to fill 60 bottles with 5 mL each. The samples were sent in labeled pairs to the individual laboratories by regular mail, encapsulated within a padded box. To ensure that the effects of mailing are minimal, a few samples have been returned after measuring and measured again to ensure their stability during transport. The two samples sent to each participant were requested to be measured in adherence of the following conditions: 1.) samples should be measured undiluted as delivered over a range of $0.1 \text{ nm}^{-1} \leq q \leq 3.0 \text{ nm}^{-1}$, 2.) at least the water background should be subtracted, 3.) if possible, the intensity should be provided in absolute units, and 4.) if possible, with uncertainty estimates of the intensity.

**Results and Discussion**

**Overview on the SAXS Measurements**

The procedures for performing SAXS measurements varied greatly between participants. Likewise, a wide spectrum of data correction procedures,[13] from very basic to very advanced, were employed by the participants. No time-correlation effect was observed in the samples for the duration of the comparison. The samples, which contain 14 wt.% of poly(acrylic acid) as stabilizer, furthermore are highly resistant to synchrotron radiation. It should be noted that silver particles with 4 wt.% stabilizer were used in an earlier attempt to perform this inter-laboratory comparison, but aggregated rapidly when irradiated with synchrotron radiation (on the order of $10^{12}$ photons s$^{-1}$). That first attempt showed a clear time-dependent drift of the incoming SAXS data three months into the inter-laboratory comparison, which led to its abortion. The particles' radiation stability will be discussed in a separate publication. The received, preprocessed data of the second batch, i.e. the stable particles discussed below, show a high degree of similarity when plotted on a double-logarithmic scale. This is evident from the overlay of the curves in Figure 1. Only two obvious outliers can be distinguished by eye.

**SAXS Data Evaluation**

The participants of this inter-laboratory study provided background-subtracted scattering curves without data evaluation. We performed a standard evaluation of the received data sets for quantification of the measurands of interest, which are 1.) the mean radius, 2.) the width of size distribution, and 3.) the particle concentration. We determined the size distributions, assuming dilute, non-interacting spherical particles of non-uniform size. Since numerous approaches exist, we chose typical representatives of three fundamentally different evaluation methods for determination of the measurands: i.) an Indirect Fourier Transformation (IFT),[16] ii.) a model fit of spheres[17] with a log-normal size distribution and iii.) a Monte-Carlo determination of size distributions.[18] Other methods such as developed by Sen[19] or usage of the mature evaluation package IRENA[20] are also suitable, but an exhaustive comparison of all available data evaluation methods and packages is not in the scope of this study. Note that the anonymized datasets are made available under a Creative Commons license for further scrutiny by interested parties (see Supporting Information data sets).

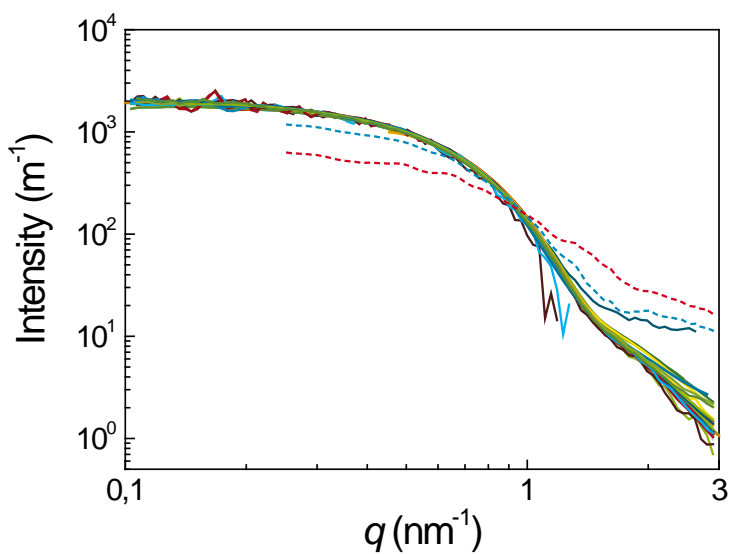

Figure 1. Overlay of 45 SAXS curves of silver particles as a function of the scattering vector, which were provided by the participants. The dashed curves are the only outliers of the study.

**Mean radius and size distribution as derived using IFT**

Here, we consider the mutual agreement between the results obtained from the different data sets. To the best of our knowledge, O. Glatter provided the first approach to determine intensity, volume and number weighted particle distributions.[16, 21] This approach was presented in 1980, and is still frequently used by many SAXS laboratories.[22] We used the IFT method for calculation of the number- and volume-weighted radii distributions, by applying a standardized analysis procedure (provided in Supporting Information). Examples of number- and volume-weighted distributions are shown in the upper part of Figure 2, as red squares and blue circles, respectively. The distributions are slightly asymmetric around their maxima, with the tail decaying more slowly towards larger radii. Therefore, symmetric functions such as a Gaussian profile cannot be considered for their description, but the log-normal function describes the distributions sufficiently well. The choice of a log-normal distribution is, furthermore, supported by theoretical considerations,[23] a transmission electron microscopy inter-laboratory study on nominally 30 nm NIST gold nanoparticles,[10] and is recommended for the standardization of the classification of magnetic nanoparticle systems.[24]

Here we employed the log-normal distribution of the radii, $R$, defined as

$$f(R) = \frac{A}{\sqrt{2\pi}\, w\, R}\, \mathrm{Exp}\left[-\frac{(\ln(R/R_0))^2}{2\, w^2}\right]. \tag{1}$$

With $A$ the area of the size distribution, $w$ the scale parameter defining the width of the size distribution, and $R_0$ the median radius, which is the value of the radii in the limit of $w = 0$. The mean value for the radii of the log normal distribution is defined by $R_0\, e^{\frac{1}{2}w^2}$ and its standard deviation $\sigma_{LogNomal} = R_0\left(e^{2w^2} - e^{w^2}\right)^{1/2}$. Examples of curve fits are shown in the upper part of Figure 2 (red and blue lines, respectively). The results of the number-weighted mean radii, $R_{n,IFT}$, and mean widths, $\sigma_{n,IFT}$, are depicted in panel (b) of Figure 2 (triangles and squares, respectively). The mean values of the data sets are $R_{n,IFT} = 2.82 \pm 0.04$ nm and $\sigma_{n,IFT} = 0.67 \pm 0.02$ nm, indicated as horizontal lines in panel (b) of Figure 2. The null hypothesis is that the data is distributed to a Student's t-distribution. This hypothesis is not rejected at the 0.05 level for $R_{n,IFT}$ and for $\sigma_{n,IFT}$, and we can thus consider the $R_{n,IFT}$ and $\sigma_{n,IFT}$ values to be normally distributed.

For a more detailed depiction of the data, the box plot in panel (c) of Figure 2 shows the distribution of $R_{n,IFT}$ and $\sigma_{n,IFT}$. It displays that 90% of the values for the radii are within the range $2.81 \leq R_{n,IFT}(\mathrm{nm}) \leq 2.83$ and the widths are within $0.67 \leq \sigma_{n,IFT}(\mathrm{nm}) \leq 0.68$. Therefore, the spread of the radii on a 90% interval is within 0.1 nm. This is surprisingly low given the relative broadness of the distribution of our particles of around 20%, in particular when compared to typical proteins or monodisperse latex particles.[11]

We repeated the IFT data evaluation procedure for the determination of the volume-weighted radii, and found mean values of $R_{v,IFT}$ = 3.22 ± 0.04 nm and $\sigma_{v,IFT}$ = 0.71 ± 0.05 nm (Figure 2 (d)). The box plots in panel (e) show that 90% of the values for the radii are within the range 3.20 nm ≤ $R_{v,IFT}$ ≤ 3.23 nm, and the widths are within 0.70 nm ≤ $\sigma_{n,IFT}$ ≤ 0.73 nm. Again, the spread of the values on a 90% interval is within 0.1 nm. The volume-weighted radii are significantly larger than the number-weighted due to the broadness of the size distribution (for monodisperse size distributions, $R_n = R_v$).

It is known that SAXS can provide precise radii if the particle size distribution is narrow, i.e. if the width of the particle size distribution can be neglected.[25] A small-angle neutron scattering round robin test on 77 nm large latex particles with a very narrow size distribution was published in 2013.[11] They found that the spread in the fitted mean particle size was about ± 1%, but the uncertainties in determination of the size distribution were much larger and sensitive to a number of instrumental effects. It is remarkable that a similarly high precision in the radius determination can be achieved also for nanoparticles with a broader size distribution (with a width of about 20%, Figure 2). As a result, we conclude that the IFT evaluation is ostensibly insensitive to the (small) variations between 1) the participants' datasets, and 2) their instruments. However, the IFT method does impose a smoothness constraint on the resulting size distribution, which may artificially constrict the results and thereby introduce an overestimated degree of precision. In the next step we therefore investigate the influence of the choice of data evaluation procedure on the results.

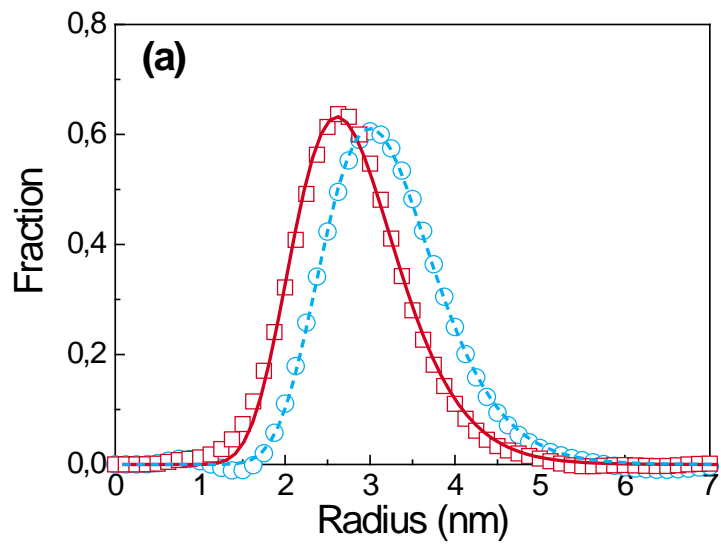

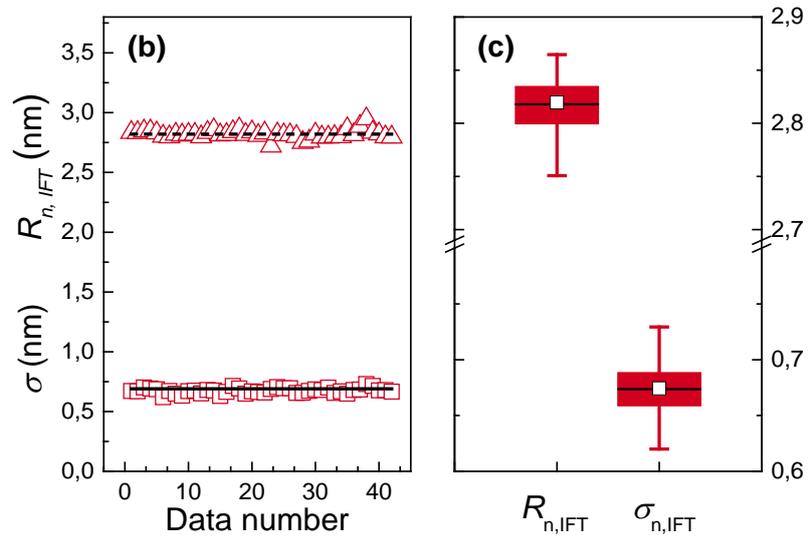

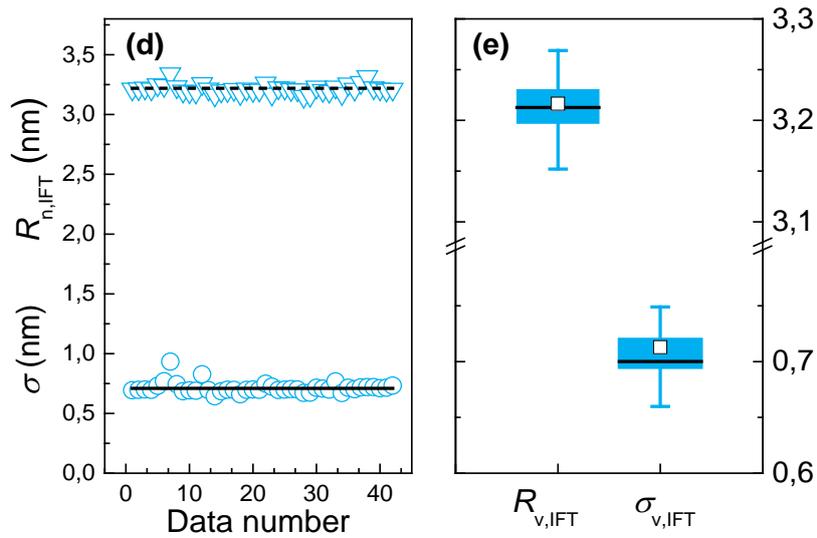

Figure 2. Results of data evaluation using the IFT method.[16] Upper row (a): Volume- and number-weighted radii distribution derived from data set number 2. Middle row (b): Number-weighted radii, $R_{n,IFT}$, and widths of the size distribution, $\sigma_{n,IFT}$, as a function of the data set number (triangles and squares, respectively). Mean values of the data sets $R_{n,IFT} = 2.82 \pm 0.04$ nm and $\sigma_{n,IFT} = 0.67 \pm 0.02$ nm are shown as horizontal lines. (c) Box plot depicts the distribution of $R_{n,IFT}$ and $\sigma_{n,IFT}$ from the measurements. The horizontal line that forms the top of the box is the 75[th] percentile ($Q_1$). The horizontal line that forms the bottom is the 25[th] percentile ($Q_3$). The horizontal line within the box is the median value and the square corresponds to the mean value. The whiskers represent lower 5% and 95% values. Bottom row (d): Volume-weighted radii, $R_{v,IFT}$, and widths of the size distribution, $\sigma_{v,IFT}$, as a function of the data set number (triangles and spheres, respectively). Mean values of the data sets $R_{v,IFT} = 3.22 \pm 0.04$ nm and $\sigma_{v,IFT} = 0.71 \pm 0.05$ nm are shown as horizontal lines. (d) Box plot of the distribution of $R_{v,IFT}$ and $\sigma_{v,IFT}$ from the measurements. Results are summarized in Table SI 2)

**Comparison of IFT with representatives of other methods**

We used SASfit[26] as a representative of a classical curve fitting procedure, and McSAS[18] as a Monte Carlo fitting program (a minimal assumption method) for SAXS data, to contrast with the aforementioned IFT results. The standardized evaluation procedures for SASfit and McSAS are described in the Supporting Information. The results obtained from both for the radii and widths are visually summarized in the curves and box plots of Figure 3 and Figure 4, respectively. All values are listed in (Table SI 2). Note that SASfit provides only estimates of number-weighted size distributions in its current implementation, and does not provide volume-weighted distributions.[26] We have chosen the log-normal distribution in SASfit for the stated reasons (vide supra).

Figure 3 and Figure 4 show that the means of the radii and widths are similar for the three different evaluation methods (means are indicated by white squares in the box plots). In order to test whether the mean values resulting from IFT, SASfit and McSAS method are the same we employed an analysis of variance (ANOVA). This demonstrates firstly that the number-weighted mean means of $R_{n,IFT}$, $R_{n,SASfit}$ and $R_{n,McSAS}$ are not equal at the 0.05 level (data mean of $R_{n,IFT}$, $R_{n,SASfit}$ and $R_{n,McSAS}$ is 2.76 nm). Secondly, the volume-weighted mean means of $R_{v,IFT}$ and $R_{v,McSAS}$ (3.22 nm and 3.18 nm, respectively, with a mean of 3.20 nm), also differ significantly at the 0.05 level. Thirdly, we found that the number-weighted mean widths $\sigma_{n,IFT}$, $\sigma_{n,SASfit}$ and $\sigma_{n,McSAS}$ are significantly different (data mean is 0.65 nm). Lastly, however, the volume-weighted mean widths of $\sigma_{v,IFT}$ and $\sigma_{v,McSAS}$ are equal at the 0.05 level (data mean is 0.71 nm). The ANOVA analysis proves that the $R_n$, $R_v$, and $\sigma_n$ are dependent on the type of evaluation method we used in this study. In contrast, $\sigma_v$ is (perhaps by chance) independent on the choice of the method.

Of interest is that the spread of the $R_n$, $R_v$, $\sigma_n$ and $\sigma_v$ values are somewhat smaller for IFT and SASfit in comparison to McSAS (see Figure 3 and Figure 4). An overview of their interquartile ranges is given in Table SI 2, where it can be seen that they are 0.03 nm (IFT), 0.02 nm to 0.03 nm (SASfit) and 0.04 nm to 0.08 nm (McSAS). The primary cause for this difference is likely the increased number of assumptions (restrictions) applied in the IFT and SASfit methods.

The values of the interquartile ranges are for all three methods small enough to recommend all three methods for data evaluation. The highly consistent results of the IFT method indicate that it is the most suited method for this particular kind of problem. The relatively wide interquartile ranges of McSAS result from its form-free nature, i.e. McSAS makes no assumption on the type, modality or smoothness of the size distribution. Therefore, we recommend a preferential use of one of the programs depending on the prior knowledge of the particles system under investigation. The IFT should be the first choice if it is known that the particles' size distribution is smooth, while McSAS is the first choice if little *a priori* knowledge is available. For example, multimodal size distributions can be detected easily with McSAS, as has been demonstrated for the reference material ERM-FD-102 (a suspension of bimodal silica particles).[26] The use of SASfit is recommended if an estimate on the size distribution is known, since it provides more than 20 different size distributions.[18] In ambiguous situations we recommend to compare the results from the different methods to verify the results.

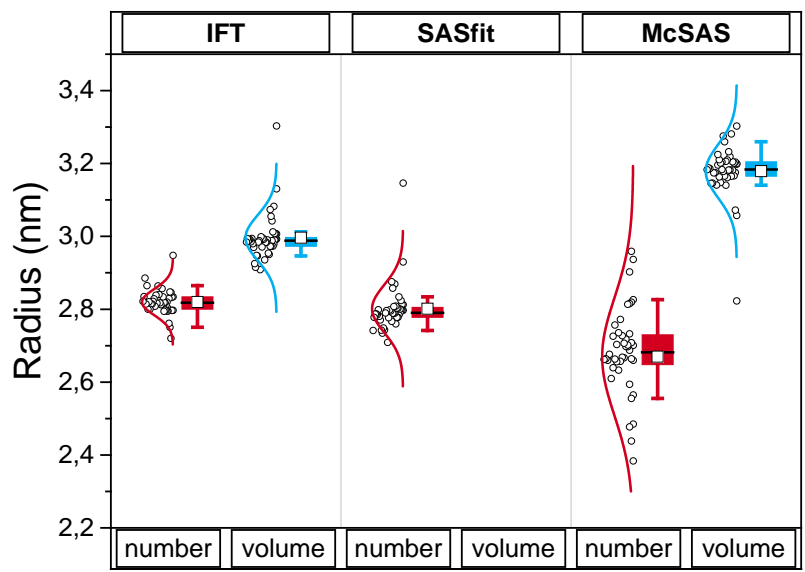

Figure 3. Comparison of number- and volume-weighted radii derived from IFT, SASfit and McSAS programs. Number-weighted values are in red, volume-weighed in blue. The top and bottom of the box delineates the 75$^{th}$ (Q$_1$) and 25$^{th}$ (Q3) percentiles. The horizontal line within the filled box is the median value and the square represents the mean value. The whiskers corresponds to lower 5% and 95% limits. The data points of the different participants are marked by spherical and quadratic symbols (number- and volume-weighted, respectively). The corresponding size distributions are displayed by solid lines. Data are listed in Table SI 2)

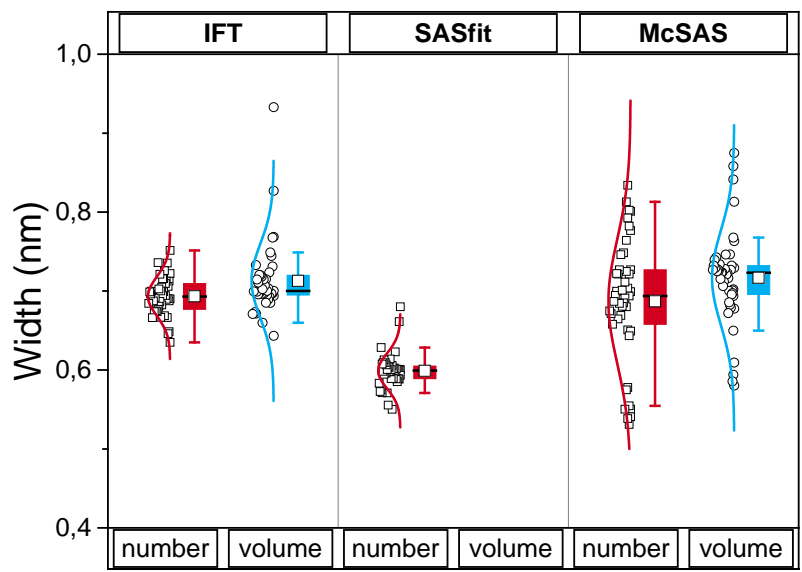

Figure 4. Comparison of number- and volume-weighted width of the radii distributions derived from IFT, SASfit and McSAS programs. Number-weighted values are in red, volume-weighed in blue. The top and bottom of the box delineates the 75$^{th}$ ($Q_1$) and 25$^{th}$ (Q3) percentiles. The horizontal line within the filled box is the median value and the square symbol is the mean value. The whiskers represent the 5% and 95% confidence interval. The data points of the different participants are marked by spherical and quadratic symbols (number- and volume-weighted data, respectively). The corresponding size distributions are displayed by solid lines. Data are given in Table SI 2.

**Accuracy and precision limits of the particle size distribution**

The estimation of the precision and accuracy of nanoparticle size distributions, referring to the closeness of agreement and the distance to the true values, respectively, is inherently challenging for a wide range of nanoscale sizing techniques. These problems arise because the outcome of particle sizing of these dimensions is generally method-specific, as discussed in a post hoc inter-laboratory comparison by Bustos et. al.[8] In this context, small-angle X-ray and small-angle neutron scattering have clear benefits of being fully traceable methods, based on first principle physics, and are capable to measure in situ size distributions of nanoparticles in the full nanoscale range of 1 to 100 nm. In principle, then, we should be able to achieve precise and accurate results.

While this work mainly details the inter-instrument variability of the findings, it is good to contrast this with the ultimately achievable accuracy and precision for a given instrument. For the determination of radii and their distributions, this means we are sensitive to variations in $q$. We have, therefore, evaluated the worst-case precision and accuracy limits of $q$ for our own instrument (an Anton Paar SAXSess). This evaluation, supplied in full in the SI as a modifiable Jupyter Notebook, is based on both considerations of the geometrical contributors to uncertainty (beam divergence, beam width, beam height, pixel or bin width, and polychromaticity), as well as the practically determinable accuracy using three different calibrants.

For our instrument, the evaluation is complicated by the use of de-smearing, which partially compensates for some of the geometrical smearing contributors. In particular, it might compensate for the worst offender: the divergence-induced spread in $q$, which is rather large in these slit-focused systems. Barring that, the binning introduces the second-worst uncertainty contribution in $q$, introducing an uncertainty of maximally 3.5% of its value (full width). Evaluating the effect of this worst-case shift in $q$ on the McSAS-retrieved distribution demonstrates that a systematic binning-induced $q$-uncertainty shift can affect the found distribution means and widths by -1/+2%, and -8/+6%, respectively.

Practical calibrants, in particular Apoferritin, showed a possible practical uncertainty in $q$ of +/- 0.035 nm$^{-1}$, which can maximally affect the found distribution means and widths by 2.5% and 35%, respectively. It was demonstrated that our instrument accuracy is well within expected limits, and therefore we have high confidence in the absolute radii values.

Uncertainties in the datapoint $q$-values are typically neglected due to their small magnitude. In summary, however, our estimates show that the uncertainty in the deduced nanostructural dimensions of the nanoparticle dispersions are affected by this. The magnitude of the effect of the uncertainty in $q$ on these dimensions, approaches the spread in the results found in this round robin experiment. We therefore strongly recommend the community to start considering uncertainty in $q$ in order to improve intercomparability and achieve ultimate nanometrological precision.

**Particle Concentration**

The particle concentration can be determined from SAXS data if the scattering intensities are provided on an absolute scale.[27] This can be achieved using water[28] or glassy carbon[29] as primary or secondary absolute calibration intensity standards. Upon the provision of data scaled to absolute units, SASfit[26] provides an estimate of particle number concentrations, which can be converted to a particle mass concentration. McSAS[18] provides estimates of volume fractions, which can be directly converted to mass concentrations. The IFT method[30] does not return any measure of particle concentration.

The intensities are given in units of $[I_{abs}(q)] = m^{-1}$, and the scattering length density difference between particles and solvent in units of $[\Delta\rho] = Å^{-2}$. For silver in water $[\Delta\rho] = 6.8 \times 10^{-5} Å^{-2}$, as calculated for an energy of 8 keV with the help of SASfit's scattering length density calculator tool (although specific contrast values were calculated and used for the different energies used by some of the laboratories). 28 data sets were provided in absolute units (labeled red in Table SI 1), and the resultant volume concentrations multiplied with the bulk density of silver of 10.49 g cm$^{-3}$ to attain mass concentration estimates of the silver nanoparticles. The number-weighted concentrations from SASfit and volume-weighted concentrations from McSAS are summarized in Figure 5 and Table SI 3. The mean number-weighted concentration was $c_{n,SASfit} = 4.20 \pm 0.73 \times 10^{-6}$ mol L$^{-1}$ and the mean volume-weighted concentration was $c_{v,McSAS} = 2.86 \pm 0.31$ g L$^{-1}$. Conversion of the number concentration to volume concentration results in $c_{v,SASfit} = 3.00 \pm 0.38$ g L$^{-1}$. An ANOVA test shows that the $c_{v,McSAS}$ and $c_{v,SASfit}$ means are not significantly different at the 0.05 level. The conversion of the volume-weighted concentration $c_{v,McSAS}$ to the corresponding number-weighted distribution results in $c_{n,McSAS} = 3.37 \pm 0.37 \times 10^{-6}$ mol L$^{-1}$. An ANOVA test shows that the means of $c_{n,SASfit}$ and $c_{n,McSAS}$ are significantly different. This demonstrates that, while it is possible to convert the number-weighted concentrations to volume-weighted ones, it is in general not recommended to convert the volume-weighted concentrations to number-weighted ones due to the divergence of the numerical nature of this operation. This has been discussed elsewhere.[18]

Both methods deliver mutually consistent values for the particle concentration, and are equally useful for this challenge. Other methods, such as ICP-MS, determine only the total silver content, from which the particle concentrations are derived based on assumptions. Therefore, it is useful to conclude that quantification of the concentration of nanoparticles with SAXS can be done straightforwardly within an uncertainty of approximately 10%.

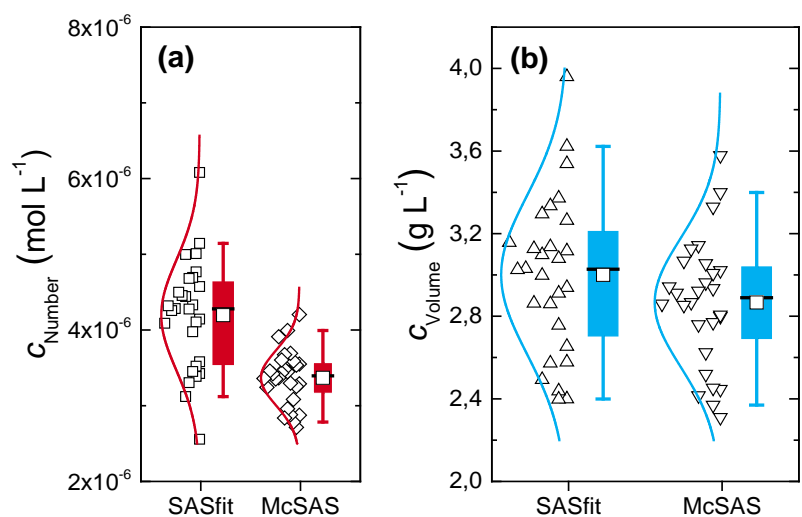

Figure 5. (a) Particle number concentration from SASfit and McSAS, with the latter converted from the volume concentration. (b) Particle mass concentration as converted from the SASfit number concentration in (a) and as a direct determination from McSAS. The white squares and horizontal lines in the box charts are the mean and median values, respectively. The lower and upper values of the box represent the quartiles $Q_1$ and $Q_3$, the upper and lower whiskers are the 5% and 95% levels. All values are summarized in Table SI 3.

**Conclusion**

Our inter-laboratory comparison demonstrates that SAXS is a mature method for particle size analysis: accurate and precise nanoparticle sizes and size distributions can be measured irrespective of the type of instrument used, be they 0.6 or 60 m in length. SAXS reliably delivers the concentration as well as the size distribution parameters with a sub-nanometer precision. We were able to confirm that SAXS is a suitable, laboratory-independent reference method for in situ nanoparticle analysis, reinforcing our opinion that SAXS is an appropriate technique for standardization and regulatory purposes regarding nanoparticle size analysis. This conclusion holds at least for monomodally distributed particles in suspension, but we expect a similar outcome for multimodal distributions or embedded nano-objects (a test to be performed in the future).

In our opinion, the cardinal benefit of SAXS is that it is inherently traceable to theoretical foundations, and that its theory is solidly grounded in first principles of physics as derived in the early stages by Debye,[31] and completed by Guinier[32], Fournet[33] and Porod[27]. This is further supported by the release of ISO 17867 on particle size analysis using SAXS.[34] From the viewpoint of regulation and validation of the technique, this is considered an important milestone for its general acceptance. It is our hope that the conclusions of this round-robin study can further serve to reinforce the elevated position of SAXS in nanoscience, regardless of the pedigree of the underlying instrument.

**Acknowledgement.**

We thank Wojciech Szczerba for a thorough discussion of the manuscript. We acknowledge all the SAXS scientists and companies who measured our samples with their instrument. For their gracious assistance with measuring the second round-robin samples, we thank: Andras Wacha, Olivier Taché, Grégory Stoclet, Frédéric de Geuser, Martha Brennich, Lauren Fullmer, Javier Pérez, Maximilian Ebisch, Jan Ilavsky, Frederick Beyer, Steven Weigand, Tilman Grünewald, Gary Bryant, Albrecht Petzold, Kushol Gupta, Andy Smith, Heinz Amenitsch, Jonathan Almer, Pierre Panine, Karsten Joensen, Jens Wenzel Andreasen, and Juan David Londono. For their gracious assistance with measuring the first round-robin samples, we thank: Jan Ilavsky, Maximilian Ebisch, Olivier Taché, Javier Pérez, Andy Smith, Otto Glatter, Albrecht Petzold, Oskar Paris, Gerhard Popovski, Linda Brützel, Michael Krumrey, Christian Gollwitzer, Frédéric de Geuser, Martin Uhlig, Regine von Klitzing, Kazuki Ito, and Pierre Panine.